\documentclass[twoside,slac_one]{revtex4}
\usepackage{graphicx}
\usepackage{fancyhdr}
\usepackage{amsmath} 
\usepackage{bm}
\usepackage{amsxtra}
\usepackage{amssymb}
\usepackage{amsthm}
\usepackage{latexsym}
\usepackage{lscape}

\begin{document}

\title{Studies of B-Hadron Decays to Charmless Final States at LHCb}

%

\author{P. Sail on behalf of the LHCb collaboration}
\affiliation{Department of Physics and Astronomy, University of Glasgow, Kelvin Building, University Avenue, Glasgow, G12 8QQ, United Kingdom}

\begin{abstract}
LHCb has a rich programme studying hadronic $B$-decays and is already showing that with early data 
it can produce precision measurements which improve on current world averages. 
The preliminary results presented include the new world's best measurements of the $A_{\it{CP}}(B_{d} \rightarrow K^{+} \pi^{-}) = -0.088~\pm~0.011~(\rm{stat})~\pm~0.008~(\rm{syst})$ and $\tau_{B_{s} \rightarrow K^{+}K^{-}} = 1.440~\pm~0.096~(\rm{stat})~\pm~0.010~(\rm{syst})~ps$. 
These were obtained using the full $2010$ data sample containing $37 \rm{pb^{-1}}$ and also $320 \rm{pb^{-1}}$ of the 2011 data. The first observation of CP violation in $B_{s} \rightarrow \pi^{+} K^{-}$ is also presented.

\end{abstract}

\maketitle

\thispagestyle{fancy}


\section{Introduction}
\label{sec:Intro}
The Large Hadron Collider Beauty (LHCb) Experiment ~\cite{ref:DetPaper} at the CERN Large Hadron Collider (LHC) is a single arm, forward facing spectrometer designed to study heavy flavour physics. The detector covers the pseudorapidity region $2 < \eta < 5$ in which the $pp$ collisions are currently at the energy of $\sqrt{s} = 7~\rm{TeV}$. The $b\bar{b}$ production cross section is $75.3~\pm~5.4~\pm~13.0~\rm{\mu b}$~\cite{ref:XSec} in the acceptance, with the collisions producing a full spectrum of $B$-hadrons (e.g. $B_{d}$, $B_{s}$, $\Lambda_{b}$). \\
\indent The study of $B_{d,s} \rightarrow h^{+} h^{'-}$ modes at LHCb relies heavily on the specific hardware and software components of the Trigger, VErtex LOcator (VELO) and Ring Imaging CHernkov (RICH) subdetectors~\cite{ref:DetPaper}.\\*
Due to the large total cross section of the $p\bar{p}$ collisions, an efficient trigger is required in order to reduce the background. The LHCb trigger comprises of a Level 0 (L0) hardware trigger, followed by a software based Higher Level Trigger (HLT) which fulfil this criterion. \\*
\indent The alignment and resolution of the VELO means it has an excellent ability to reconstruct tracks accurately which is important for proper time resolution measurements etc. There are two RICH sub-detectors that are part of the LHCb, combined these cover the momentum range $1-100~GeV/c$. It is these which give excellent Particle IDentification (PID) used to separate hadronic final states (e.g $\pi, K, p$). \\*
\indent Presented here are measurements of the $B_{d} \rightarrow K^{+} \pi^{-}$ and $B_{s} \rightarrow \pi^{+} K^{-}$ CP asymmetries (section ~\ref{subsec:Acp}) and effective lifetime of $B_{s} \rightarrow K^{+}K^{-}$ (section ~\ref{subsec:Lifetime_top}) performed using integrated luminosities of $320~\rm{pb^{-1}}$ (2011 data) and $37~\rm{pb^{-1}}$ (full 2010 data) respectively.
 
\section{Analysis of $B_{d,s} \rightarrow h^{+} h^{'-}$ at LHCb}
\label{sec:Analysis}

The family of $B_{d,s} \rightarrow h^{+}h^{'-}$ modes, where the discussion here will focus primarily on $B^{0}_{d}$ or $B^{0}_{s}$ decaying to a final state $\pi$ or $K$, are of particular interest when studying Charge Parity (CP) violation in neutral $B$ meson mixing or in their decay. These decay modes are dominated by loop processes with a small contribution from tree processes. The former are sensitive to Beyond the Standard Model (BSM) effects influencing the decay processes. The presence of these BSM effects can be observed by measuring quantities associated with decay modes which follow these paths, such as $B_{s} \rightarrow K^{+} K^{-}$. These can be used to determine if there are any deviations between the observed quantities and Standard Model (SM) predictions. Asymmetry measurements of flavour specific final states can also be measured in order to observe any deviations from the CP violation predicted by the SM.

\subsection{Time integrated CP Asymmetry of $B_{d} \rightarrow K^{+} \pi^{-}$ and $B_{s} \rightarrow \pi^{+} K^{-}$ }
\label{subsec:Acp}
The time-integrated CP asymmetry, $A_{\it{CP}}$, is defined as 

   \begin{equation}
     A_{\it{CP}} = \frac{\Gamma(\bar{B}\rightarrow\bar{f}) - 
       \Gamma(B\rightarrow f)}{\Gamma(\bar{B}\rightarrow\bar{f}) + \Gamma(B\rightarrow f)}
       \label{eq:}
   \end{equation}
   
\noindent for a given final state $f$. In addition to comparing the value to SM predictions, the measurement of $A_{\it{CP}}$ for $B_{d} \rightarrow K^{+} \pi^{-}$ and $B_{s} \rightarrow \pi^{+} K^{-}$ is also an approximate test of U-spin symmetry where the $\rm d \leftrightarrow s$ (see Equation~\ref{eq:USpin}) quarks interchange.

\begin{align}
	\nonumber  A_{\it{CP}}(B_{d} \rightarrow K^{+} \pi^{-}) \approx A_{dir}(B_{s} \rightarrow K^{+} K^{-}) \\
	A_{\it{CP}}(B_{s} \rightarrow \pi^{+} K^{-})\approx A_{dir}(B_{d} \rightarrow \pi^{+} \pi^{-})
	\label{eq:USpin}
\end{align}

The asymmetry measurements of the flavour specific, time integrated modes; $B_{d} \rightarrow K^{+} \pi^{-}$ and $B_{s} \rightarrow \pi^{+} K^{-}$ discussed here use $\approx 320 ~\rm{pb^{-1}}$ of data collected in 2011~\cite{ref:B2HH2011Data}. These measurements are updated from a previous analysis using data collected in 2010 ~\cite{ref:B2HH2010Data}. There are many overlapping $B_{d,s} \rightarrow h^{+} h^{'-}$ final states that are kinematically very similar. The use of RICH PID information is imperative to discriminate between these specific backgrounds, such as $B_{d} \rightarrow \pi^{+} \pi^{-}$ and $B_{s} \rightarrow K^{+} K^{-}$. 

For a final state $f$, the physical asymmetry $A_{\it{CP}}$ is related to the raw asymmetry $A_{\it{CP}}^{RAW}$ by

 \begin{equation}
		A_{\it{CP}}^{RAW}(B_{d,s} \rightarrow f) = A_{\it{CP}}(B_{d,s} \rightarrow f) + A_{D}(B_{d,s} \rightarrow f) + A_{P}(B_{d,s}) .
\end{equation}  

\begin{figure}[htbp]
  \begin{center}
    \includegraphics*[width=0.85\textwidth]{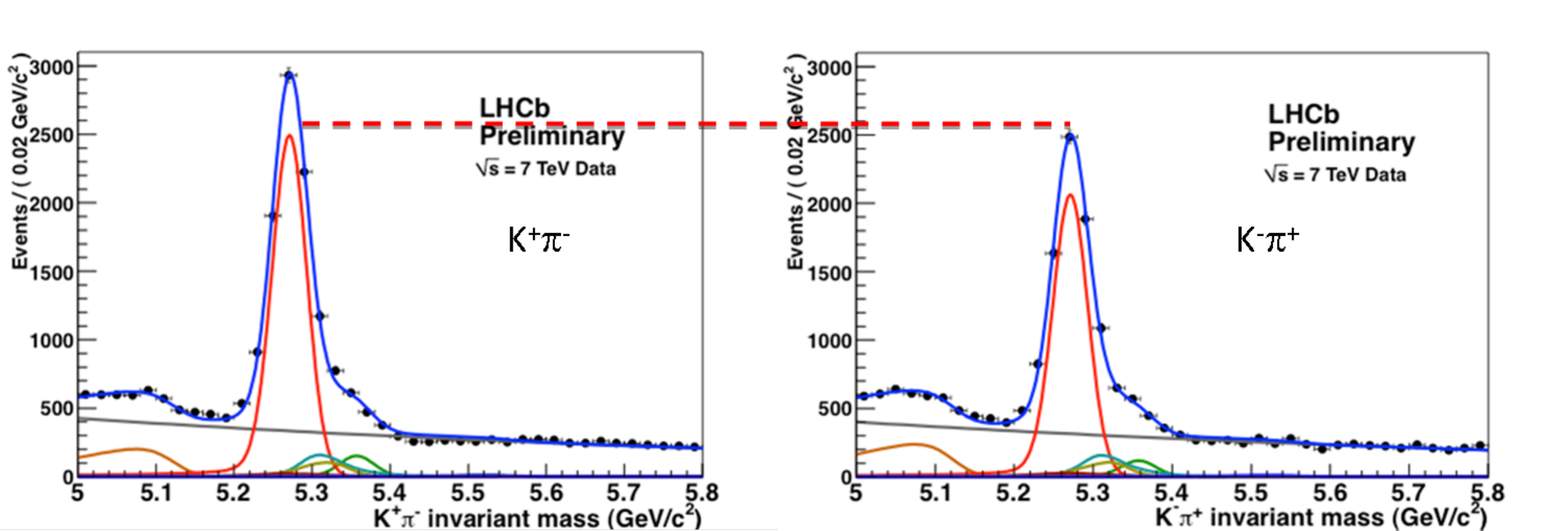}
  \end{center}
  \caption{Raw asymmetry between $B_{d} \rightarrow K^{+} \pi^{-}$ and $\bar{B_{d}} \rightarrow K^{-} \pi^{+}$ ~\cite{ref:B2HH2011Data}. The event selection is tuned to maximise the sensitivity to $A_{\it{CP}}$ for $B_{d} \rightarrow K \pi$ events.}
  \label{fig:bd2kpiRawAsymmetry}
\end{figure}

\begin{figure}[htbp]
  \begin{center}
    \includegraphics*[width=0.85\textwidth]{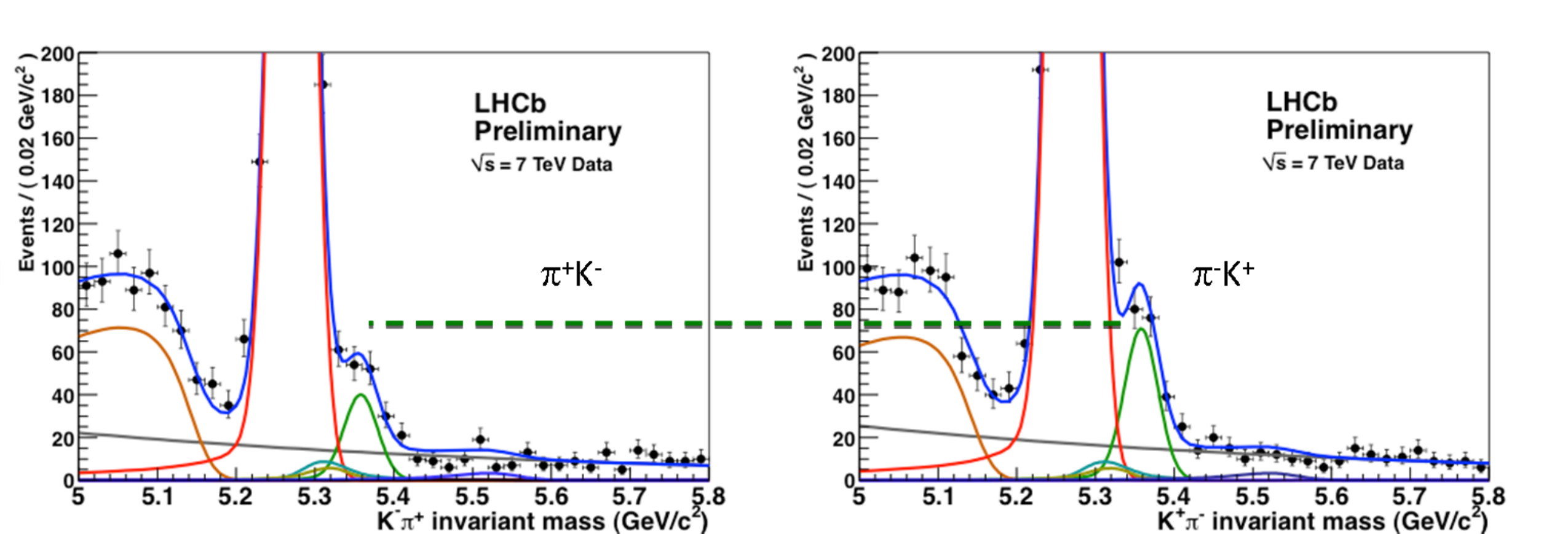}
  \end{center}
  \caption{Raw asymmetry between $B_{s} \rightarrow \pi^{+}K^{-}$ and $\bar{B_{s}} \rightarrow \pi^{-}K^{+}$ ~\cite{ref:B2HH2011Data}. The event selection is tuned to maximise the sensitivity to $A_{\it{CP}}$ for $B_{s} \rightarrow \pi K$ events.}
  \label{fig:bs2KpiRawAsymmetry}
\end{figure}

The raw asymmetry between the untagged final states of $B_{d} \rightarrow K \pi$ (Figure~\ref{fig:bd2kpiRawAsymmetry}) and $B_{s} \rightarrow \pi K$ (Figure~\ref{fig:bs2KpiRawAsymmetry}) can be observed by extracting the time integrated signal yields before applying the following corrections. 

The two corrections which need to be applied to extract the physical asymmetry are:

\begin{itemize}
	 \item $A_{D}$: detector-induced $K^{+}\pi^{-}/K^{-}\pi^{+}$ charge asymmetries
	 \item $A_{B}$: B production asymmetry
\end{itemize}

The detector-induced asymmetry, $A_{D}$, is determined by using untagged $D^{0}$ and self-tagged $D^{* \pm} \rightarrow D^{0} \pi_{s}^{\pm}$ decays in high statistics charm samples. The slow pion, $\pi_{s}^{\pm}$, allows for the tagging of the $D^{*\pm}$. From these, the raw asymmetries of $D^{0}$ to $KK$, $\pi\pi$ and $K\pi$ final states can be measured. This is done by employing the current world average for the $\it{CP}$ asymmetries for the two modes $D^{0} \rightarrow K^{+}K^{-}$ and $D^{0} \rightarrow \pi^{+} \pi^{-}$, with the $\it{CP}$ asymmetry for the Cabibbo-favoured $D^{0} \rightarrow K^{-} \pi^{+}$ assumed to be negligible. This reduces the number of unknown observables to four: $A_{D}(\pi_{s})$, $A_{D}(K\pi)$, $A_{P}(D^{*})$, and $A_{P}(D^{0})$. Considering the equations in~\ref{eq:DetAsym}, these can then be solved, to acquire the relevant detector asymmetry, $A_{D}(K\pi)$.

\begin{align}
	\nonumber  A_{\it{CP}}^{RAW }(K\pi)^{*} &= A_{\it{CP}}(K\pi) + A_{D}(\pi_{s}) + A_{D}(K\pi) + A_{P}(D^{*}) \\
	\nonumber  A_{\it{CP}} ^{RAW}(KK)^{*} &=  A_{\it{CP}}(KK) + A_{D}(\pi_{s}) + A_{P}(D^{*}) \\
	\nonumber  A_{\it{CP}}^{RAW}(\pi\pi)^{*} &= A_{\it{CP}}(\pi\pi) + A_{D}(\pi_{s}) + A_{P}(D^{*}) \\
	A_{\it{CP}}^{RAW}(K\pi) &=  A_{\it{CP}}(K\pi) + A_{D}(K\pi) + A_{P}(D^{0}) 
	\label{eq:DetAsym}
\end{align}


As the LHC is a p-p collider, and the initial state is not $\it{CP}$-symmetric, it is possible that the production of $B_{d,s}$ and $\bar{B}_{d,s}$ mesons will occur at different rates within the acceptance of the detector. 
An estimate of this production asymmetry is determined by using the self-tagged mode $B^{0} \rightarrow J/\psi(\mu^{+}\mu^{-})K^{*0}(K\pi)$, where the flavour of the $B$ meson is tagged via the $K^{*0}$ final states such that $K^{*0} \rightarrow K^{+} \pi^{-}$ or $\bar{K}^{*0} \rightarrow K^{-} \pi^{+}$ ~\cite{ref:B2HH2011Data}. As a result of the additional corrections, the physical asymmetries become

\begin{align}
	A_{\it{CP}}(B_{d} \rightarrow K^{+} \pi^{-}) &= -0.088~\pm~0.011~(\rm{stat})~\pm~0.008~(\rm{syst})~and  \\
	A_{\it{CP}}(B_{s} \rightarrow \pi^{+} K^{-}) &= 0.27~\pm~0.08~(\rm{stat})~\pm~0.02~(\rm{syst})
	\label{eq:PhysicalAsymm}
\end{align}

\noindent which are consistent with results from the data collected in 2010~\cite{ref:B2HH2010Data}.

\subsection{Lifetime measurements of $B_{s} \rightarrow K^{+} K^{-}$}
\label{subsec:Lifetime_top}

The $B_{s}$ decay to $K^{+}K^{-}$ is a CP even final state accessible from both mass eigenstates; $|B_L\rangle$ (light) and $|B_H\rangle$ (heavy). In the absence of $\it{CP}$ violation it would only be accessible from the $|B_L\rangle$ state, however the SM predicts a very small amount of CP violation which allows a contribution from $|B_H\rangle$. A larger than expected contribution to the decay from $|B_H\rangle$ would be an indication of CP violation beyond what is expected from the SM.
The lifetime distribution is a function of two exponential functions where the corresponding untagged rate can be written as


\begin{align}
     \langle \Gamma (B_{s}(t) \rightarrow f) \rangle &= R_{H}^{\it{f}} e^{-\Gamma_{H}^{{(s)}} t} + R_{L}^{\it{f}} e^{- \Gamma_{L}^{(s)} t} \\
     & \propto (1+A_{\Delta \Gamma}) e^{-\Gamma_{H}^{(s)}t} + (1-A_{\Delta \Gamma}) e^{-\Gamma_{L}^{(s)}},
\end{align}

\noindent where $A_{\Delta\Gamma}$ is defined as:

\begin{center} 
	$A_{\Delta\Gamma} = \frac{R_{H} - R_{L}}{R_{H} + R_{L}}$ \hspace{0.25cm} with the predicted value by the SM of \hspace{0.25cm}
	$A_{\Delta\Gamma}(B_{s} \rightarrow K^{+}K^{-}) = -0.97_{-0.009}^{+0.014}$~\cite{ref:EffectiveBsLifetime}	
\end{center}

The dominant experimental challenge associated with measuring the lifetime of this decay, is due to the introduction of a lifetime bias in the selection of these events. In order to account for this effect, two independent methods were used ~\cite{ref:Bs2KK2010Lifetime}; a relative and an absolute lifetime measurement. These methods were both optimised for the full 2010 data set of $\approx 37~\rm{pb^{-1}}$. The details of each will be discussed in the following subsections.

\subsubsection{Relative lifetime measurement}
\label{subsubsec:Relative}

The relative lifetime method for measuring the $B_{s} \rightarrow K^{+}K^{-}$ lifetime uses the assumption that the lifetime acceptance introduced via the event selection, can be cancelled by using a kinematically similar channel as the acceptances are expected to be similar. The channel chosen is the relatively abundant and similar $B_{d} \rightarrow K^{+} \pi^{-}$. 
The sample of $B$ meson candidates obtained from the event selection is split into thirty proper-time bins split between  $-1~\rm{ps}$ and $35~\rm{ps}$, with the binning chosen such that each bin is expected to contain approximately the same number of B meson candidates. \\
\indent The $B_{s} \rightarrow K^{+}K^{-}$, $B_{d} \rightarrow K^{+}\pi^{-}$ and $B_{s} \rightarrow \pi^{+}K^{-}$ mass distributions are all described by Gaussian functions with first order polynomials used for the background. The parameters of the signal and background PDFs are fixed to the results of the time-integrated mass fits before the lifetime fit is performed. The $B_{d} \rightarrow K^{+}\pi^{-}$ yield ($N_{B \rightarrow K \pi}$) is constrained to follow equation ~\ref{eq:RelEq}, where $\bar{t_{i}}$ is the mean proper time in the $\it{i^{\rm{th}}}$ bin. A simultaneous fit is then performed to the $K^{+}K^{-}$ and $K^{\pm}\pi^{\mp}$ invariant mass spectra over all bins. The difference in reciprocal lifetimes, $\tau_{B_{s} \rightarrow K^{+}K^{-}}^{-1} - \tau_{B_{d} \rightarrow K \pi}^{-1}$ (Equation ~\ref{eq:DeltaKK}), is extracted directly from the fit.

\begin{center}
  \begin{equation}
	N_{B_{s}\rightarrow K^{+}K^{-}}(\bar{t_{i}}) = N_{B_{d} \rightarrow K \pi}(\bar{t_{i}})R_{0} \exp^{ -\bar{t_{i}} \Delta_{ K^{+}K^{-} } }
    \label{eq:RelEq}
  \end{equation}
\end{center}

\noindent where $\Delta_{K^{+}K^{-}}$ is defined as

\begin{center}
	\begin{equation}
		\Delta_{K^{+}K^{-}} \equiv \Gamma_{B_{s} \rightarrow K^{+}K^{-}} - \Gamma_{B_{d} \rightarrow K\pi}
		\label{eq:DeltaKK}
	 \end{equation}
\end{center}

\begin{figure}[htb !!!!!!!!!]
	\begin{center}
     		 \includegraphics[width=0.4\textwidth]{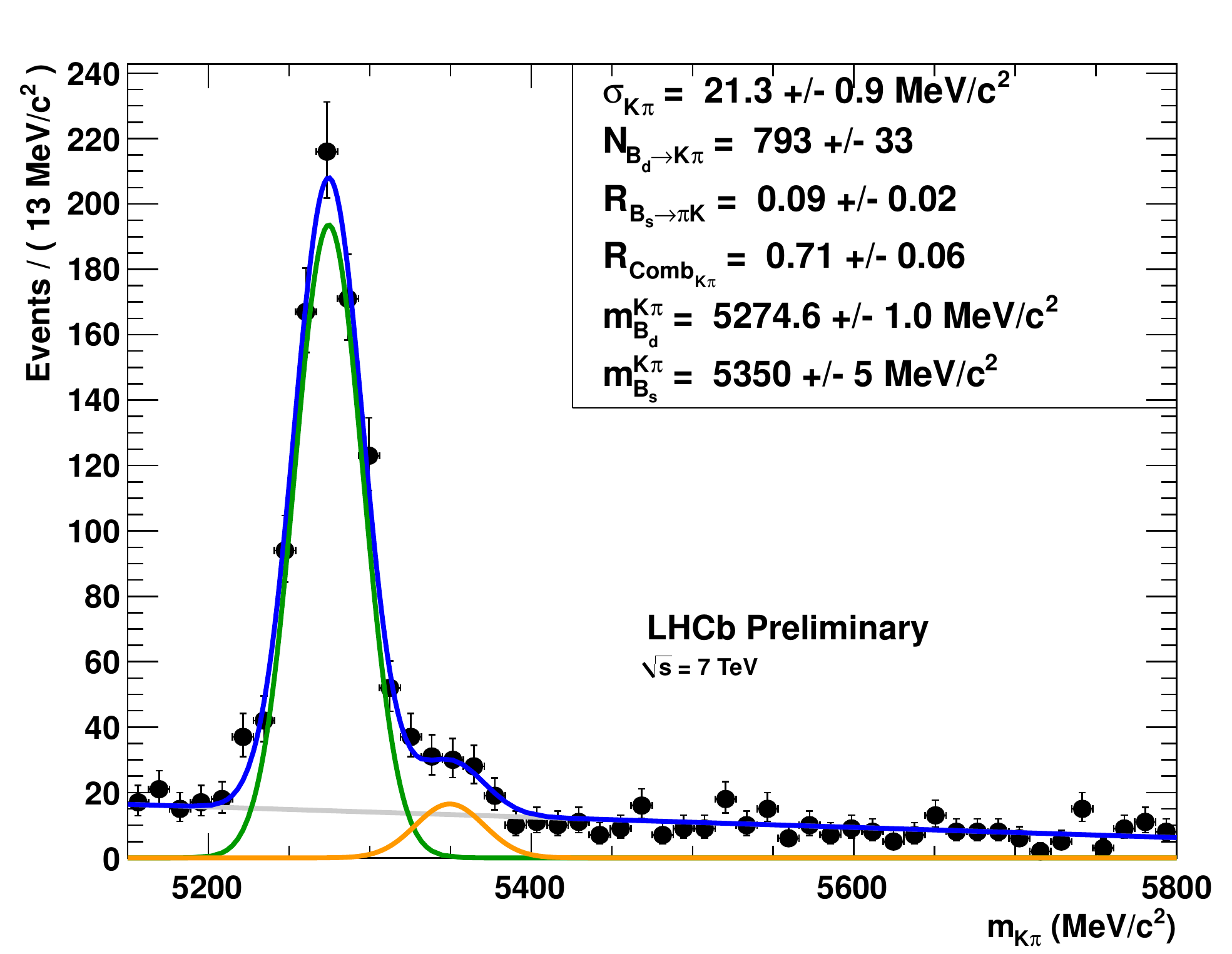} 
     		 \includegraphics[width=0.4\textwidth]{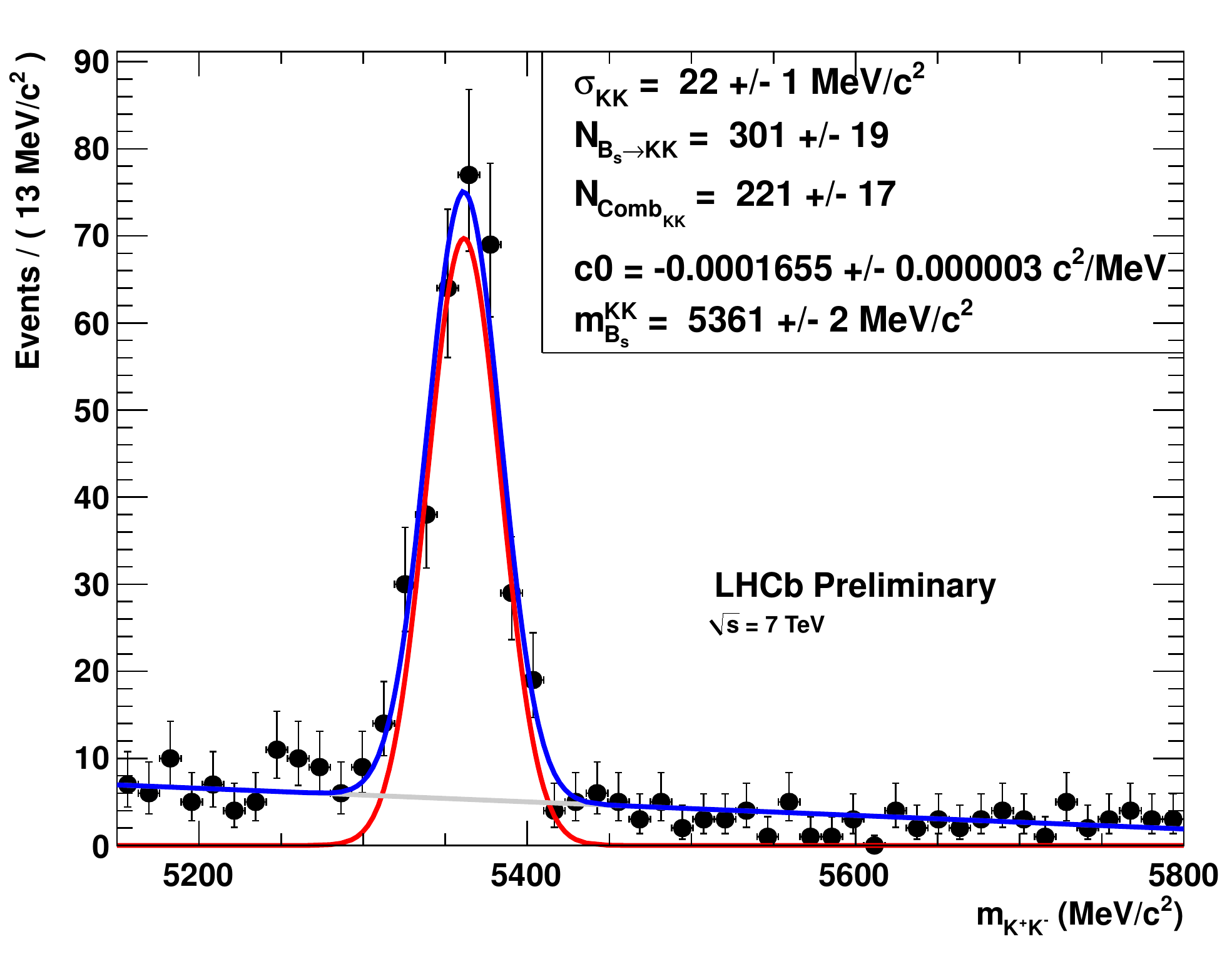}
     		 \caption{Invariant mass distributions for kinematically similar $B_{d} \rightarrow K^{+} \pi^{-}$ (left) and $B_{s} \rightarrow K^{+} K^{-}$ using relative lifetime optimised cuts ~\cite{ref:Bs2KK2010Lifetime}.} 
		\label{fig:RelMass}
	\end{center}
\end{figure}

The fitted quantity of the relative lifetime measurement is thus dependent on the world average of $\tau_{B_{d}}$ as an input, to obtain $\tau_{B_{s} \rightarrow K^{+} K^{-}}$. 


\subsubsection{Absolute lifetime measurement}   
\label{subsubsec:Absolute}

\begin{figure}[!!!htbp]
  \centering
   \begin{minipage}{0.32\textwidth}
    (a)\\
    \includegraphics[width=0.9\textwidth]{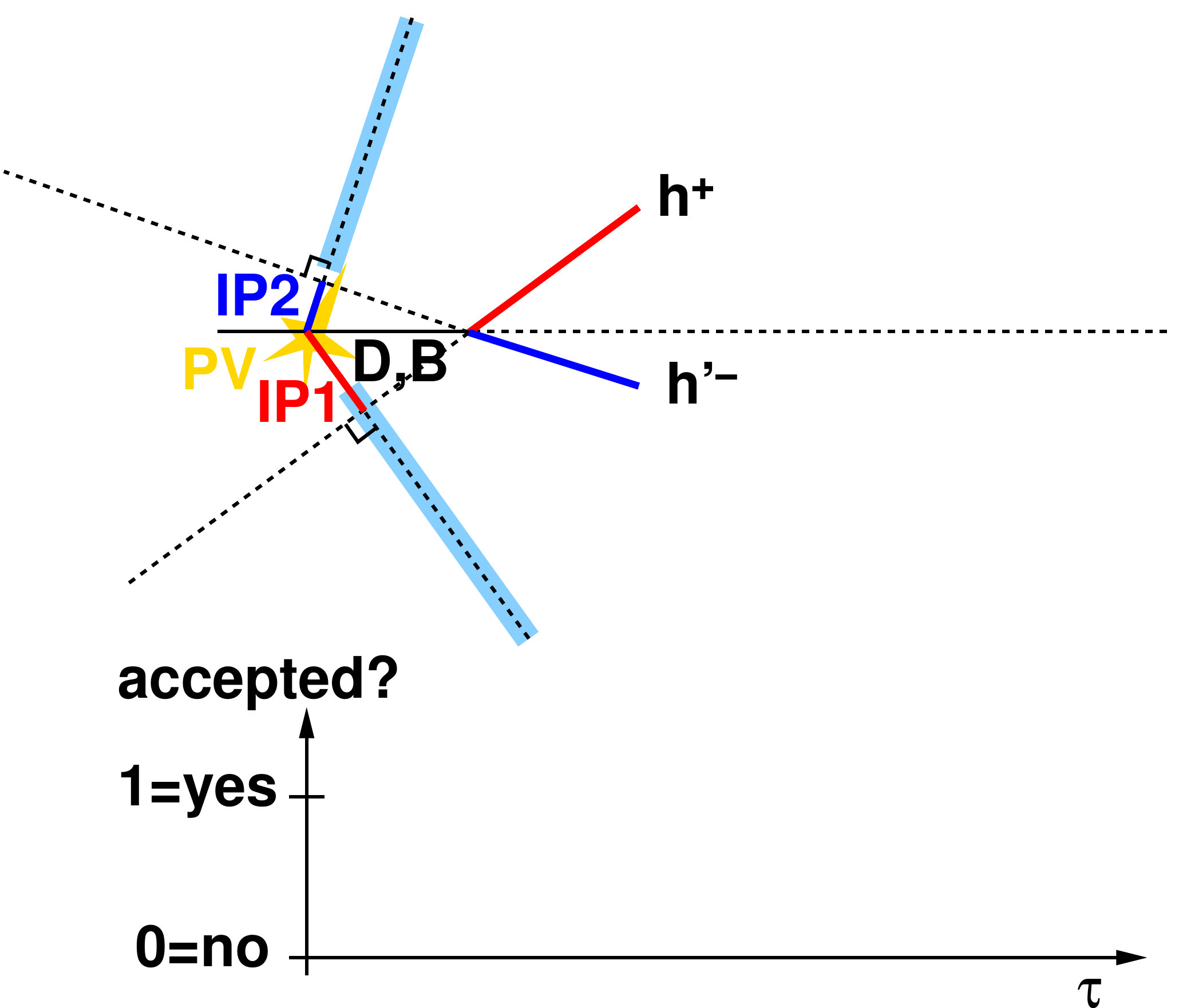}
  \end{minipage}
  \begin{minipage}{0.32\textwidth}
    (b)\\
    \includegraphics[width=0.87\textwidth]{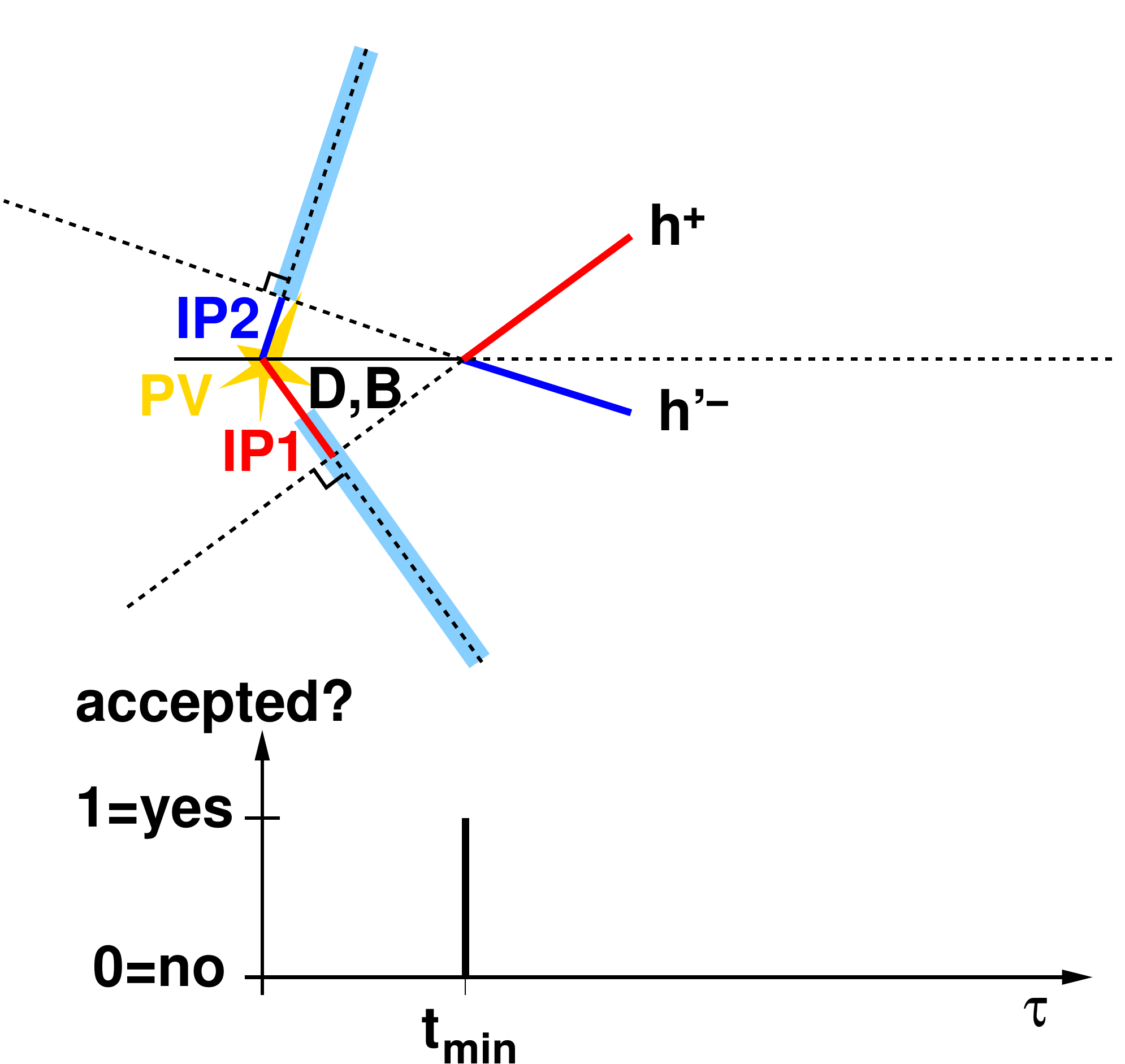}
  \end{minipage}
  \begin{minipage}{0.32\textwidth}
    (c)\\
    \includegraphics[width=0.81\textwidth]{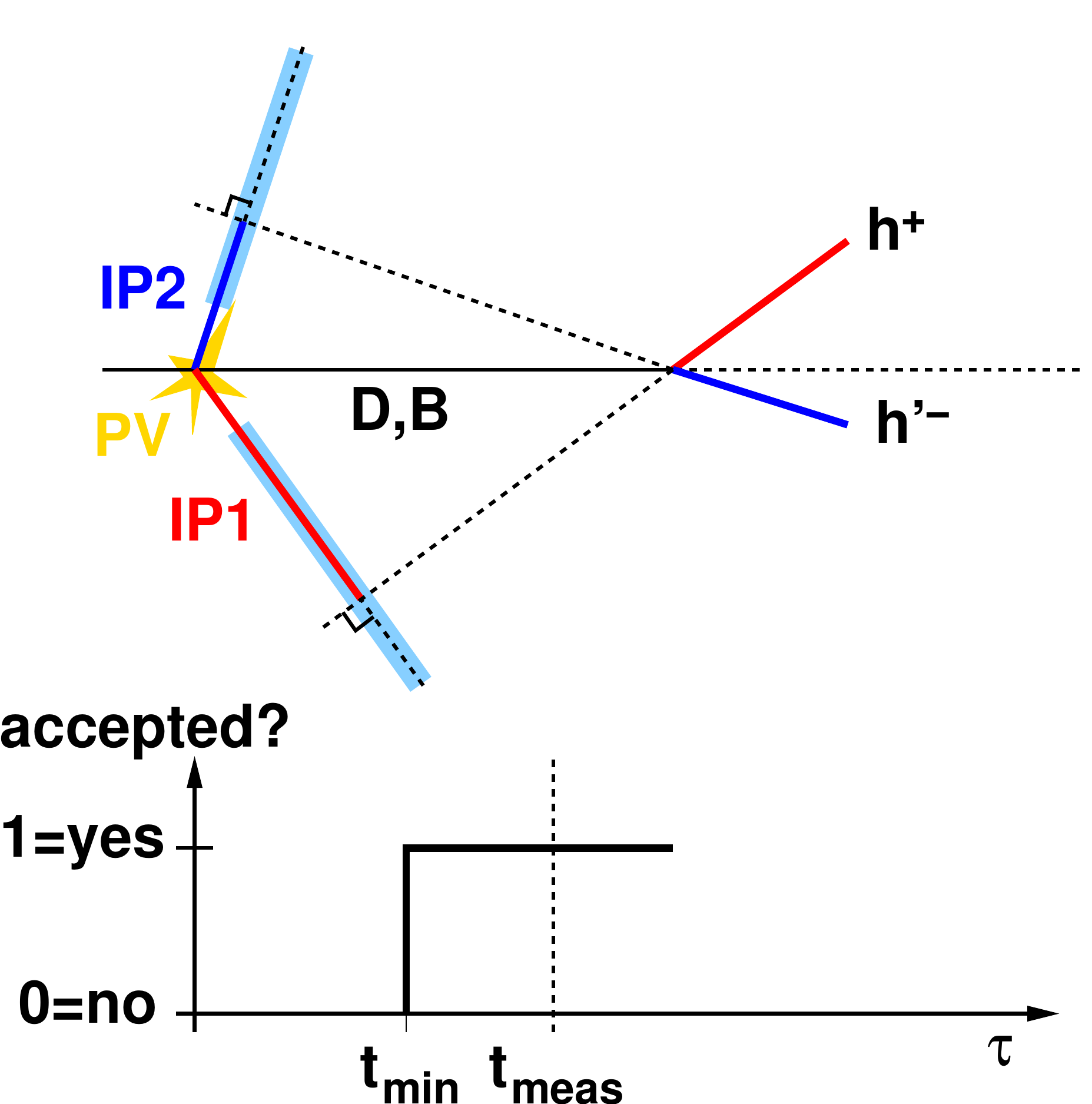}
  \end{minipage}
  \vspace{10mm}\\
  \caption[Lifetime acceptance function for an event with a two-body
    hadronic decay.]{Proper-time acceptance function for an event of a
    two-body hadronic decay. The light blue (shaded) regions show the
    bands for accepting the impact parameter of a track. The impact
    parameter is too small in (a) it reaches the accepted range in
    (b). The actual measured lifetime lies in the accepted region
    (c). The acceptance intervals give conditional likelihoods used in
    the lifetime fit ~\cite{ref:Bs2KK2010Lifetime}.}
  \label{fig:AcceptanceCorrect}
\end{figure}

The absolute lifetime measurement measures the $B_{s} \rightarrow K^{+} K^{-}$ lifetime value directly using a novel approach to determine the acceptance effect ~\cite{ref:CERNSwim1, ref:CDFSwim1, ref:CDFSwim2, ref:LHCbSwim1, ref:LHCbSwim2, ref:Bs2KK2010Lifetime} that is introduced via the event selection. The acceptance correction is calculated by taking selected events and moving the Primary Vertices (PV) of each event along the flight direction of the $B$ meson, which is determined by its momentum vector. The trigger and event selection criteria are then rerun, all lifetime dependent parameters are re-evaluated for the new position, and a lifetime interval in which the event would be accepted is determined. In the case of the absolute lifetime measurement, only the event selection is rerun as this selection was made tighter than the trigger cuts. The event by event acceptance intervals are then used as parameters in the fit. Figure ~\ref{fig:AcceptanceCorrect} displays this method schematically. \\



\begin{figure}
\begin{center}
      \includegraphics[width=0.49\textwidth]{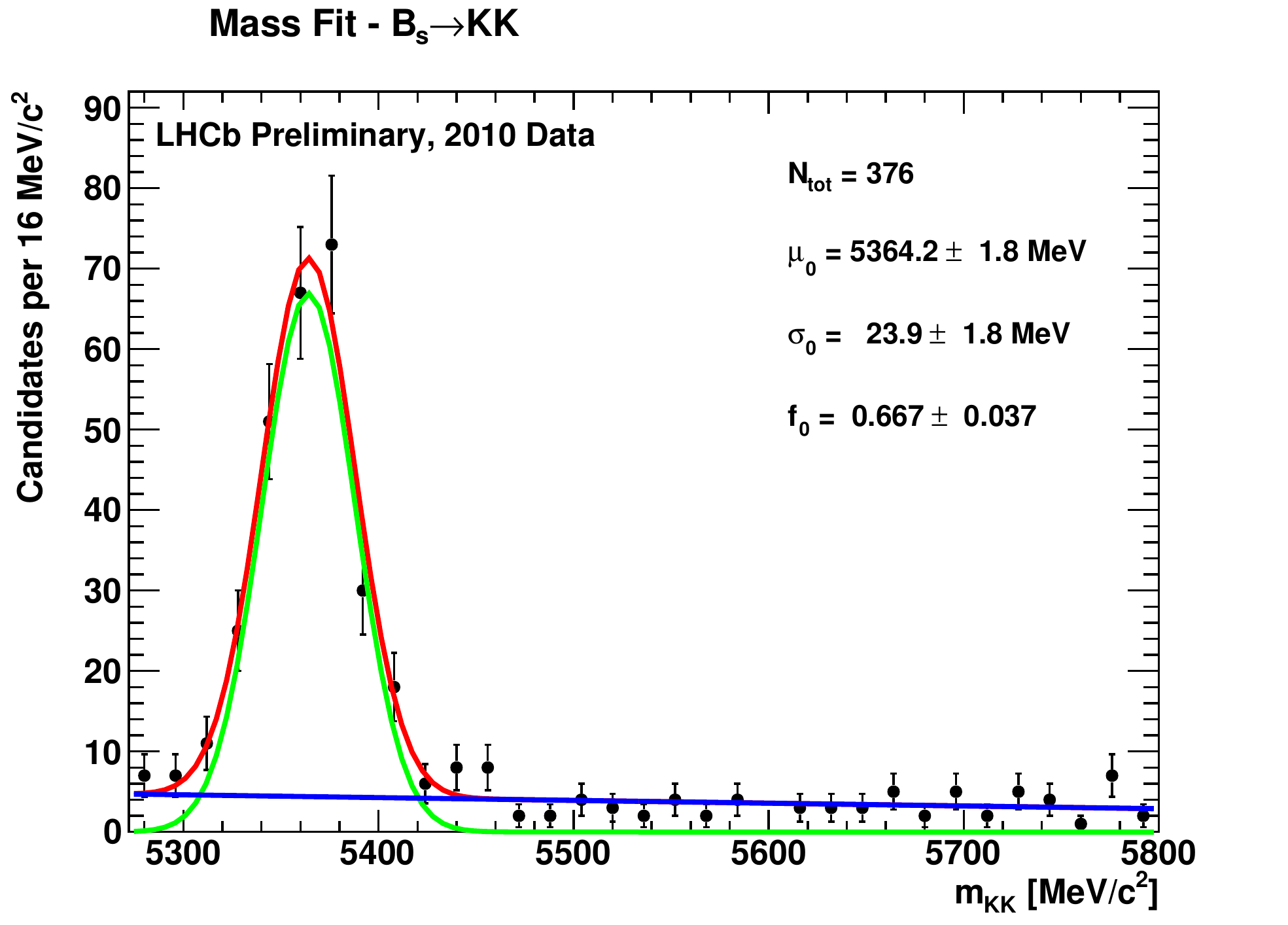}
      \includegraphics[width=0.49\textwidth]{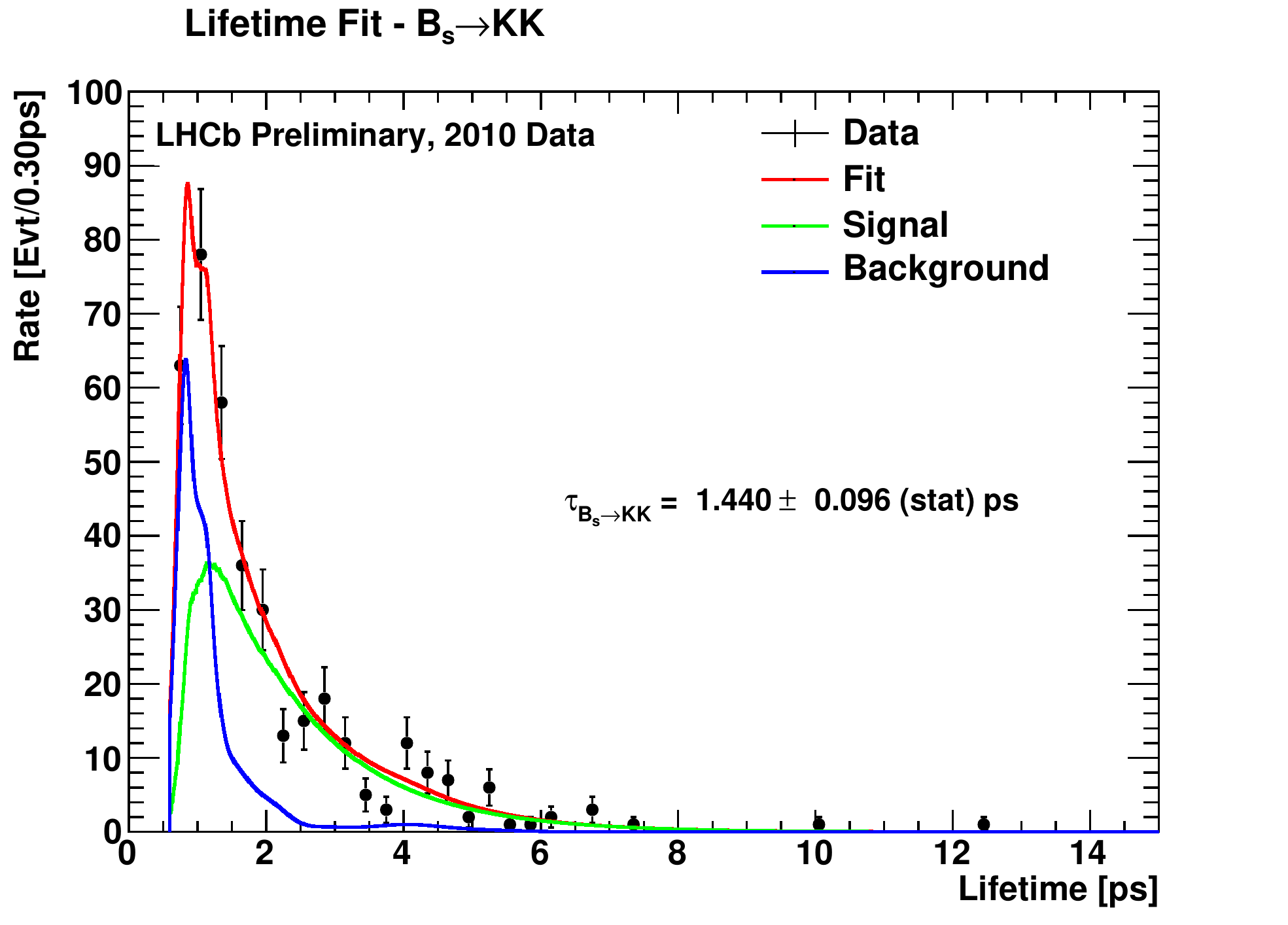}
      \caption{The mass and lifetime distribution of $B_{s} \rightarrow K^{+} K^{-}$ candidate events. The signal fraction of the data set is extracted (left) and then used to confirm the signal and background composition for the unbinned maximum likelihood fit (right) ~\cite{ref:Bs2KK2010Lifetime}.}
      \label{fig:AbsLife}
\end{center}
\end{figure}

The effective $B_{s} \rightarrow K^{+}K^{-}$ lifetime is extracted using an unbinned maximum likelihood fit, which use both an analytical Probability Density Function (PDF) for the signal lifetime and a non-parametric estimated PDF for the combinatorial background. The measurement is factorised into two components. \\
\indent The first is a fit to determine a signal and background probability per event, this is applied to the observed invariant mass spectrum. To reduce the impact of partially reconstructed background and $B_{d}^{0}$ decays interfering with the mass spectrum fit, the mass range is reduced to $5272-5800~\rm{MeV/c^{2}}$. The signal is fitted using a standard Gaussian distribution with a linear distribution used to describe the combinatoric background. \\
\indent The resultant signal and background probabilities are used in the second fit, which is performed to extract the lifetime. The lifetime PDF  of the combinatorial background is estimated from data by using a non-parametric method and is modelled by a sum of kernel functions~\cite{ref:LHCbSwim2, ref:Kernel1}. These kernel functions represent each candidate event with a normalised Gaussian function which is centred around the proper time of the event. The kernel functions are also weighted by the probability of that each event belonging to that particular class. The width of these is determined by an estimate of the density of the candidates at that particular lifetime. \\
\indent The resultant signal, background and total lifetime distributions of the data before corrections can be seen in Figure ~\ref{fig:AbsLife}. 
The fitted value is in good agreement with standard model predictions and improves upon the previous world measurement produced by CDF~\cite{ref:CDFBs2KKlifetime}. 

{\small
\begin{description}
     
      \item \bf{LHCb Preliminary:} $\hat{\tau}_{B_{s} \rightarrow K^{+}K^{-}}^{Absolute} = 1.440~\pm~0.096~(\rm{stat})~\pm~0.010~(\rm{syst})~\rm{ps}$
      
      \item \bf{CDF Preliminary:} $\hat{\tau}_{B_{s} \rightarrow K^{+}K^{-}}^{CDF} = 1.58~\pm~0.18~(\rm{stat})~\pm~0.02~(\rm{syst})~\rm{ps}$
      
      \item \bf{SM Prediction:} $\hat{\tau}_{B_{s} \rightarrow K^{+}K^{-}}^{SM} = 1.390~\pm~0.032~\rm{ps}$ ~\cite{ref:EffectiveBsLifetime}
      
\end{description}
}

\subsection{Rare decay observation}

\begin{figure}[htp !!!!!!!!!!!]
\begin{center}
      \includegraphics[width=0.49\textwidth]{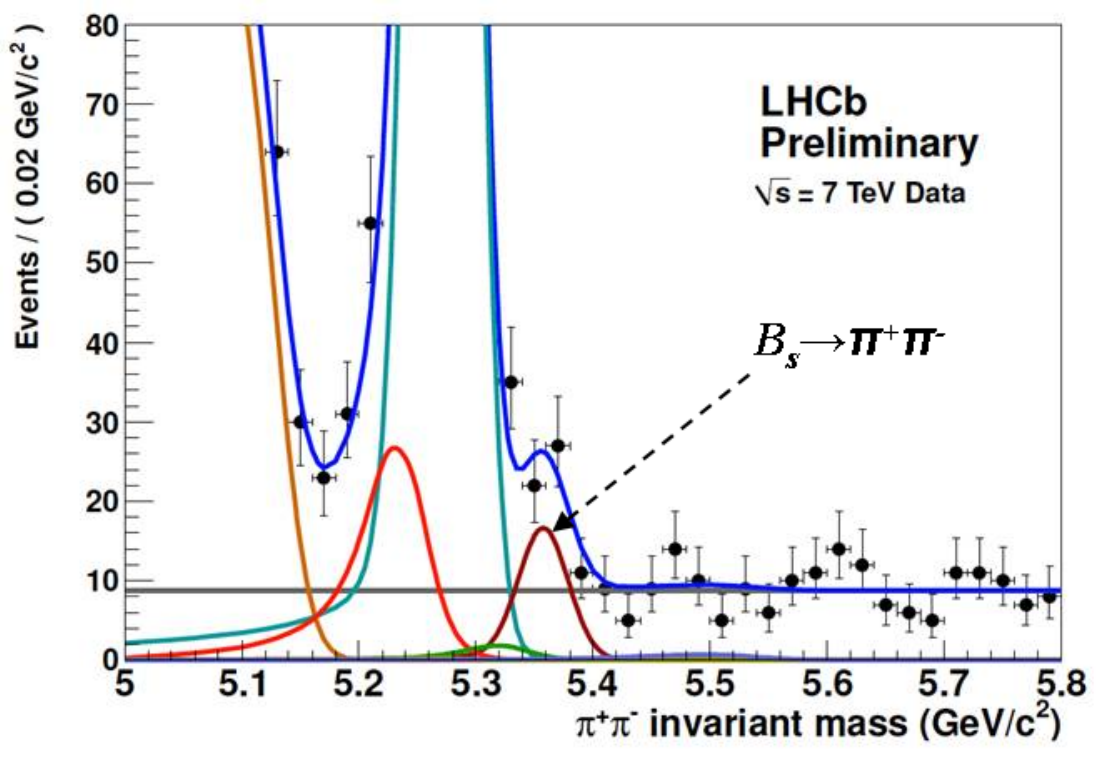}\\
      \caption{Reconstructed invariant mass distribution showing $B_{s} \rightarrow \pi^{+}\pi^{-}$ signal ~\cite{ref:B2HH2011Data}. The main signal components seen are:  
      $B_{s} \rightarrow \pi^{+}\pi^{-}$ (brown), $B_{d} \rightarrow \pi^{+}\pi^{-}$ (light blue), $B_{d} \rightarrow K \pi$ (red), $B_{s} \rightarrow \pi K$ (green), combinatorial background (grey) and 3-body partially reconstructed decays (orange).}
      \label{fig:RareDecay}
\end{center}
\end{figure}

A first observation at LHCb of the rare $B^{0} \rightarrow h^{+}h^{'-}$ mode,  $B_{s} \rightarrow \pi^{+}\pi^{-}$ (Figure ~\ref{fig:RareDecay}), has been made to a significance of $5.3\sigma$ using $\approx 320~\rm{pb^{-1}}$ of the 2011 data ~\cite{ref:B2HH2011Data}. 
The preliminary measurement of the Branching Ratio (BR) is

	\begin{center}

	  
	  	$BR( { B_{s} \rightarrow \pi^{+}\pi^{-} } ) = 0.98_{-0.19}^{+0.23}~(\rm{stat})~\pm~0.11~(\rm{syst})~\times~10^{-6}$.
	  
	  \end{center}

\noindent This value can be compared with the SM predicted value to look for inconsistencies.

\section{Summary}

LHCb has already produced many competitive measurements using the $\approx 37~\rm{pb^{-1}}$ from the 2010 data taking run at $\sqrt{s} = 7~\rm{TeV}$. The rich programme of studies in the $B^{0} \rightarrow h^{+}h^{'-}$ sector has already measured a new worlds best measurment for the $\tau_{B_{s} \rightarrow K^{+}K^{-}}$ via two independent analysis methods. This should be improved on with the larger data sample acquired during 2011-2012. We also observe the first evidence of CP violation in the $B_{s} \rightarrow \pi^{+}K^{-}$ mode to a significance of $3\sigma$, as well as improving on the current world average of $A_{CP}(B_{d} \rightarrow K^{+}\pi^{-})$ using $\approx 320~\rm{pb^{-1}}$ of 2011 data. 
Observations of rare decays in this family are also beginning to become evident, with the first observation of $B_{s} \rightarrow \pi^{+}\pi^{-}$ to $>5\sigma$ being made. \\
A data set of $\approx 1~\rm{fb^{-1}}$ is expected by the end of 2011, with which LHCb will be able to probe the limits of the SM yet further in the search for New Physics (NP) .


\bigskip 

\end{document}